\documentclass[12pt,preprint]{aastex}

\shorttitle{High-energy emissions from the Crab pulsar}
\shortauthors{Takata et al.}

\begin{document}


\title{Polarization of high-energy emissions from the Crab pulsar}


\author{J. Takata}
\affil{ASIAA/National Tsing Hua University - TIARA,
PO Box 23-141, Taipei, Taiwan}
\email{takata@tiara.sinica.edu.tw}
\author{H.-K. Chang} 
\affil{Department of Physics and Institute of Astronomy,
 National Tsing Hua University, Hsinchu 30013, Taiwan}

\and
\author{K.S. Cheng}
\affil{Department of Physics, University of Hong Kong, Pokfuam Road,
Hong Kong, China}




\begin{abstract}
We investigate polarization of  high-energy emissions from the Crab pulsar in
the frame work of the outer gap accelerator, following previous works of 
Cheng and coworkers. The recent version of the outer gap, which 
extends from inside the null charge surface to the light
cylinder, is used for examining the synchrotron 
radiations from the secondary and
the tertiary pairs, which are produced outside the gap. 
We calculate the light curve, the spectrum and the polarization
characteristics, simultaneously, by taking into account gyration motion of
the particles. The polarization position angle curve and
the polarization degree are calculated to compare with the
Crab optical data. We demonstrate that the radiations from inside 
the null charge surface make outer-wing and  off-pulse emissions in 
the light curve,  and the tertiary pairs contribute to bridge
emissions. The emissions from the secondary pairs explain the 
 main features of the observed  light curve and  spectrum.
 On the other hand, both emissions from inside the null 
charge surface and from the tertiary pairs are required to explain
the optical polarization behavior of  the Crab pulsar.  The energy
dependence of the polarization features is expected by the present
model.  For the Crab pulsar, the polarization position angle curve indicates 
that the viewing angle of the observer measured from the rotational
axis is  greater than $90^{\circ}$.

\end{abstract}





\section{INTRODUCTION}

The \textit{Compton Gamma-Ray Observatory} (CGRO) had
shown that  young pulsars are strong $\gamma$-ray sources, and
 had detected seven $\gamma$-ray pulsars (Thompson et al. 1999). The CGRO
revealed that  the light curve with double peaks in a
period and the spectrum extending to above GeV are typical 
features of the high-energy emissions from the $\gamma$-ray pulsars.  
Although these data have constrained  proposed models,
the origin of the $\gamma$-ray emission is not yet conclusive.
One important reason is that various models have
successfully explained the features of the observed spectra and/or light
curves. For example, in the frame works of the  polar cap model (Daugherty
\& Harding 1996), the two-pole caustic model (Dyks \& Rudak 2003) and 
the outer  gap model (Romani \& Yadigaroglu 1995; Cheng et al. 2000, hereafter CRZ00), the  main features 
of the observed light curve such as two peaks in a period and the
emissions between the two peaks are all expected. So, we cannot discriminate 
the three different models using  the observed light curve. 
Furthermore, both the polar cap and outer gap models have explained the
observed $\gamma$-ray spectrum (Daugherty \& Harding 1996; Romani 1996).

Polarization measurement will play an
important role to discriminate the various models, because it increases
the number of observed parameters, namely, 
polarization degree (p.d.) and position
angle (p.a.) swing. So far, only the optical polarization 
data for the Crab pulsar is available (Smith et al. 1988; Kanbach et
al. 2005) in high energy bands. For the Crab pulsar, the spectrum is
 continuously extending from optical to $\gamma$-ray bands. 
In addition, the pulse positions in the wide energy bands are all in
phase, which would indicate that  
the optical emission mechanism is related to 
higher energy emission mechanisms.  In the future, the next generation 
Compton telescope  will probably be able to measure 
polarization characteristics in MeV bands. These data
will be useful for discriminating  the different models.

Chen et al. (1996) considered the polarization characteristics in
the peaks of the light curve for the Crab pulsar with an outer gap
model. In that model, the synchrotron radiation was used. 
 The model assumed that the charge particles are distributed with a
 Gauss function in the azimuthal direction to guarantee 
the formation of the peaks in the light curve.
Romani \& Yadigaroglu (1995) calculated the polarization characteristics
predicted by the curvature radiation process in the frame work of the
one pole outer gap model. In that model, however, 
the optical polarization data of the Crab pulsar was reproduced by  very
specialized selection of the model parameters such as the inclination
angle and the viewing angle of the observer (Dyks et al. 2004, 
hereafter DHR04). DHR04 showed that  the two-pole caustic geometry, 
in which the acceleration region extends from the stellar surface to
near the light cylinder, explains the pattern of the p.d. and the fast swing
of the p.a. at both peaks in the light curve. DHR04 also showed that the
effect of depolarization due to overlap of the emissions from the 
different magnetic field lines is not strong so that 
the intrinsic level of the polarization degree at each
radiating point remains in  the bridge and off-pulse phases of the
light curve.  Recently, P\'etri \& Kirk (2005) proposed 
that the optical emission originates from outside the light cylinder
 and calculated the polarization 
characteristic predicted by the pulsar striped wind model 
(Kirk et al. 2002). However,  the observations show that the pulse peaks of 
the radio, optical, X-ray and $\gamma$-ray are all in phase, and  it
is not clear  how the pulsar striped wind can radiate in multi frequency. 
Furthermore, all of the previous models have not
considered the spectrum, the light curve and the polarization all
together. 

In this paper, we examine the optical polarization characteristics of
the Crab pulsar with the spectrum and the light curve predicted by
modifying  the 3-D outer gap model in CRZ00.  
CRZ00 has calculated the synchrotron self-inverse Compton scattering
process of the secondary pairs produced outside  the outer gap 
and has explained the Crab spectrum from X-ray to $\gamma$-ray bands. 
 Zhang \& Cheng (2002) reconsidered CRZ00 model and 
calculated the energy dependent light curves and the phase resolved
X-ray spectrum. In CRZ00, however, the outer-wing 
and the off-pulse emissions of the Crab pulsars cannot be
reproduced, because  the traditional outer gap geometry,  
which extends  from the null charge surface of the
Goldreich-Julian charge density (Goldreich \& Julian 1969) to 
the light cylinder along the magnetic field lines and about
$180^{\circ}$ in azimuthal direction, is assumed. Furthermore, 
 the spectrum in the optical band was not 
considered. In this paper, on these grounds, we modify 
the CRZ00 geometrical model into a more realistic model, following 
 recent 2-D electrodynamical studies (Takata et al. 2004, 2006; 
Hirotani 2006), and  
we examine the light curve, the spectrum and the polarization
characteristics of the Crab pulsar.  

In section~\ref{model}, we present the high-energy
emission model and the calculation method for the polarization. 
In section~\ref{result}, we compare the polarization characteristics in
the optical band with the Crab data, and demonstrate 
 that the present model reproduces
the observed light curve, the spectrum and the polarization
characteristics all together. We diagnose the viewing angle for various
inclination angle and for various emission height 
by comparing the model with the Crab optical data.  
In~section~\ref{discuss}, we 
 predict the polarization characteristics in higher energy bands.

\section{EMISSION MODEL}
\label{model}
The outline of the outer gap model for the Crab pulsar is as follows. 
The charge particles are accelerated 
by the electric field (section~\ref{oustruc}) parallel to the
magnetic field lines in  so called gap, where the charge density is
different from the Goldreich-Julina charge density (Goldreich \& Julian
1969), $\rho_{GJ}\sim -\mbox{\boldmath$\Omega$}\cdot\mbox
{\boldmath$B$}/2\pi c$, where
$\mbox{\boldmath$\Omega$}$ is the angular velocity, $\mbox{\boldmath$B$}$ is the local
magnetic field and $c$ is the speed of light.  
The high energy particles accelerated in the gap 
emit the $\gamma$-ray photons (called primary photons) via the curvature
radiation process. For the Crab pulsar, most of the primary photons
escaping from the outer gap will convert into secondary pairs 
outside the gap, where the accelerating electric field vanishes, 
by colliding with surface and/or synchrotron X-rays emitted 
by the secondary pairs. 
The secondary pairs emit  optical - MeV photons via the
synchrotron process (section~\ref{synemi}) and photons above MeV with
the inverse Compton process.  The high-energy photons emitted by the
secondary pairs may convert into tertiary pairs at higher altitude by
colliding with the soft X-ray from the stellar surface.
 The tertiary pairs emit the optical-UV
photons via the synchrotron process (section~\ref{tertiary}).
 This secondary and tertiary photons appear as the observed 
radiations from the Crab pulsar. In section~\ref{result}, we will show that 
 although the observed main features of the light curve and the spectrum are
explained by the emissions from the secondary pairs, the observed 
polarization characteristics are explained with the emissions 
from the tertiary pairs, which  were not considered in CRZ00. 

In section~\ref{oustruc}, we describe the outer gap structure. 
In section~\ref{synemi}, we discuss the synchrotron emission 
from the secondary and the tertiary pairs. The calculation method of the 
polarization are described in section~\ref{stokes}, In
section~\ref{para}, we introduce the model parameters of the present study.
 
\subsection{Outer gap structure}
\label{oustruc}
Because the Crab pulsar has a thin gap, we describe the accelerating
electric field (Cheng et al. 1986a, 1986b) with 
\begin{flushleft}
\begin{equation}
E_{||}(r)=\frac{\Omega B(r)f^2(r)R_{lc}^2}{c s(r)},
\label{electric}
\end{equation}
\end{flushleft}
where $f(r)$ is the local gap thickness in units of the light radius,
$R_{lc}=c/\Omega$, and $s(r)$ is the curvature radius of the magnetic
field line. This traditional outer gap model assumes that 
 the outer  gap starts from the null charge surface. However, it has 
been well known that the traditional outer gap geometry 
cannot reproduce the off-pulse emission of the Crab pulsar.  
On the other hand, recent 2-D electrodynamical studies 
(e.g. Takata et al. 2004; Hirotani 2006)  for the outer gap
accelerator have demonstrated that the inner boundary 
of the outer gap is shifted toward the stellar surface 
from the null charge surface by the
current through the gap. Therefore,  we take into account 
the radiation and the pair-creation processes inside the null charge
surface. In such a case, the emissions between 
the stellar surface and the null charge surface contribute to the 
light curves as the outer-wing and the off-pulse emissions. 

As demonstrated 
by the 2-D electrodynamical model, the electric field inside the  
null charge surface  rapidly decreases along the magnetic field line  
 because of the screening effects of the pairs produced 
near the inner boundary. To simulate the accelerating 
electric field inside null charge
surface in the present geometrical study,  we assume that
 the strength of the accelerating field changes quadratically 
along the magnetic field line  as 
\begin{equation}
E_{||}(r)=E_{n}\frac{(r/r_{i})^2-1}{(r_{n}/r_{i})^2-1},~~r_i\leq r\leq r_n, 
\label{electeq}
\end{equation}
where $E_{n}$ is the strength of the electric field at the null charge
surface and  $r_{n}$ and $r_{i}$ are the radial distances to the 
null charge surface and the inner boundary of the gap, respectively. 
For the inclined rotator, the radial distance to the null charge
surface varies for different field lines so that the $r_n(\phi)$ 
is a function of the azimuthal angle ($\phi$). In this paper, the ratio
$r_i(\phi)/r_n(\phi)$ is assumed to be a constant for each field line 
(that is, no azimuthal dependence of $r_i/r_n$),  and is treated  
as a model parameter. The local Lorentz
factor of the accelerated particles (called primary particles) 
in the outer gap  is described by  
$\Gamma_p(r)=[3s^2(r)E_{||}/2e]^{1/4}$ with assuming
the force balance between the acceleration and the curvature radiation
back reaction.

Rhe primary photons emitted in the outer gap 
may  make the pairs inside and outside the gap with the soft X-ray
from the stellar surface or the synchrotron radiation of the secondary pairs. 
 CHRb (1986) considered the pair creation 
process in the gap
between the primary curvature photons and the soft photons emitted by
the secondary pairs, which was produced outside of the gap by the
pair-creation process of the primary photons,  and estimated   
the typical  fractional gap size as   
$f\sim33.2B^{-13/12}_{12}P^{33/20}$, where  $B_{12}$ is the strength of the
stellar magnetic field  in
unit of $10^{12}$~G and $P$ is the rotational period. However, the
 soft photons emitted from the secondary pairs may be beamed out of
the outer gap, because the secondary pairs are created just outside
the gap and emit the photons to the convex side
of the field lines. The screening pairs will be created by the
pair-creation process  between the primary curvature photons 
and the surface X-rays. In such a case, 
Zhang \& Cheng (1997) estimated  typical fractional size of the outer
gap as  
\begin{equation}
f(R_{lc}/2)~\sim 5.5B_{12}^{-4/7}P^{26/21}.
\label{fsize}
\end{equation}
The local fractional size of the outer gap is estimated by
$f(r)\sim f(R_{lc}/2)(2r/R_{lc})^{1.5}$ (CRZ00).
In this paper, we use the fractional gap size $f(R_{lc})=0.11$ with
$B_{12}=3.7$, which is inferred from the dipole radiation model of 
the pulsar spin down. 

The present 3-D geometrical model assumes 
that the outer gap extends around the whole polar cap, 
 because we have not had any reliable model for the 3-D geometry
 of the acceleration region. If we consider the emission
region extending very close to the light cylinder $\rho\sim R_{lc}$, 
the expected light curve with a moderately changing emissivity
along the field line  may have triple or more peaks. 
Therefore, we expect that the emissivity  near the light cylinder is
declined and/or the radiations from near the light cylinder are beamed
out of  line of sight due to the  magnetic bending. In the calculation, 
we constrain the boundaries of the
axial distance and radial distance for the emission regions with
$\rho_{\max}=0.9R_{lc}$ and $r=R_{lc}$, respectively.

\subsection{Synchrotron emission from the pairs}
\label{synemi}
\subsubsection{secondary pairs}
The primary photons escaping the outer gap convert into the secondary
pairs by colliding the non-thermal X-ray photons, which were emitted
by the synchrotron process of the secondary pairs. 
The photon spectrum of the synchrotron
radiation by the secondary pairs is described by (CRZ00)
\begin{equation}
F_{syn}(E_{\gamma},r)=\frac{3^{1/2}e^3B(r)\sin\theta_{p}(r)}{mc^2h
E_{\gamma}}\int\left[\frac{dn_e(r)}{dE_e}\right]F(x)dE_edV_{rad},
\label{emis}
\end{equation}
where $x=E_{\gamma}/E_{syn}$, $E_{syn}(r)=3he\Gamma_s^2(r)
B(r)\sin\theta_p(r)/4\pi m_ec$ is the typical photon energy of the
secondary pairs, $\Gamma_s$ represents Lorentz factor of the secondary
pairs, $\theta_p$ is the pitch angle of the particle,
$F(x)=x\int_x^{\infty} K_{5/3}(y)dy$, where $K_{5/3}$ is the modified
Bessel function of order 5/3, and $dV_{rad}$ is the volume element of the
radiation region considered. The distribution of the pairs is given by 
\begin{equation}
\frac{dn_e}{dE_{e}}\sim\frac
{l_{cur}n_{GJ}\mathrm{ln}(E_{cur}/E_e)}{\dot{E}_eE_{cur}},
\label{dist}
\end{equation}
 where 
$l_{cur}=eE_{||}c$ is the local power of the curvature radiation, 
$n_{GJ}=\Omega B/2\pi ce$ is the Goldreich-Julian number density, 
$E_{cur}(r)=3h\Gamma_{p}^3(r)c/4\pi s(r)$ is the characteristic energy
of the curvature photons emitted by the primary particles
 and $\dot{E}_{e}= 2e^4B^2(r)\sin^2\theta_p(r)
\Gamma_s^2/3m^2c^3$ is the energy loss rate of the synchrotron
radiation of the secondary pairs.
 
The pitch angle of the secondary pairs is estimated from 
$\sin\theta_p(R_{lc})\sim \lambda/s(R_{lc})$, where $\lambda$ is the
mean free path of the pair-creation between the primary $\gamma$-rays
and the non-thermal X-rays from the secondary pairs. The mean free path is
estimated from $\lambda^{-1}\sim  n_X\sigma_{\gamma\gamma}$, where
$n_X$ is the typical non-thermal X-ray number density and 
$\sigma_{\gamma\gamma}$ is the pair-creation cross section, which is
approximately given by $\sigma_{\gamma\gamma}\sim \sigma_{T}/3$,
where $\sigma_{T}$ is the Thomson cross section. For the Crab pulsar,
the typical number density becomes 
$n_{X}\sim L_{X}(<E_X>)/\delta\Omega R_{lc}^2
c <E_X>\sim 8\times10^{17}~\mathrm{cm}^3$,  where we used the typical 
energy $<E_X>\sim (2m_ec^2)^2/10
~\mathrm{GeV}\sim100$~eV, the typical non-thermal X-ray luminosity $L_X\sim
10^{35}\mathrm{erg/s}$, and the solid angle $\delta\Omega=1$~radian.
 As a result, the mean free path  becomes $\lambda\sim
10^{7}$~cm so that the pitch angle is estimated by 
$\sin\theta_{p}\sim\lambda/s(R_{lc})\sim 0.06$, where we used the
curvature radius $s(R_{lc})=R_{lc}$. 
In this paper, therefore, we adopt $\sin\theta_p(R_{lc})=0.06$. The
local pitch angle is calculated from
$\sin\theta_p(r)=\sin\theta_p(R_{lc})(r/R_{lc})^{1/2}$.

The outer gap extends above the last-open lines with the
thickness $f(R_{lc})$. And then,
we assume that the secondary pair region extends just above the
outer gap with the thickness $\lambda$.
\subsubsection{tertiary pairs}
\label{tertiary}
Some high-energy photons emitted by the inverse Compton process of the
secondary pairs may convert into tertiary pairs at
higher altitude by colliding  with thermal X-ray photons from the star. 
The energy of the
new born tertiary pairs will be described by the most energetic secondary 
photons  from the secondary pairs.  
According to the study by CRZ00, the most energetic
($\sim$1GeV) secondary photons via the inverse Compton process of the
secondary pairs are about one order magnitude smaller than that of
the primary photons ($\sim$10GeV), which make the secondary
pairs. Therefore, we expect that the tertiary pairs are produced with 
a Lorentz factor of one order magnitude smaller than that 
of the secondary pairs. The optical depth of the pair-creation between
the high-energy photons emitted by the secondary pairs and the thermal
X-ray photons from the stellar surface is estimated as $\tau\sim
n_X\sigma_{\gamma\gamma}R_{lc}\sim 0.1$, with the typical thermal X-ray
number density  $n_X\sim4\pi R_*^3\sigma T^4/4\pi R_{lc}^2 c
k_BT\sim 5\times 10^{15}~\mathrm{/cm^3}$, where
$R_*=10^{6}~\mathrm{cm}$ is the stellar radius, $\sigma$ is the
Stefan-Boltzmann constant, $k_B$ is the Boltzmann constant, and $T$ is
the surface temperature, for which we adopt the reasonable 
value $T=2\cdot10^{6}$~K (Yakovlev \& Pethick 2004). 
In this paper, therefore, we use that
  the maximum energy of and  the local 
number density of the tertiary pairs are smaller than about $10\%$ of 
 those of the secondary pairs.  Because the pitch  angle of the
pairs  increases with altitude, we use $\sin\theta_p=0.1$ for the
pitch angle of the tertiary pairs. In fact, the results 
 are not sensitive to the pitch angle of the tertiary pairs.    

The tertiary pairs are produced above the region of the secondary
pairs. As we will see in section~\ref{lcurve}, the emissions from the
secondary pairs make the main features of the light curve
such as the two peaks in a period, and the emissions from the tertiary
pairs contribute to the bridge phase.

\subsubsection{synchrotron cooling}
\label{cooling}
For the Crab pulsar, the observed spectrum has a spectral break around
10~eV. This feature was not considered in  CRZ00. One possibility
of the explanation of the observed break is due to the effect of the 
synchrotron cooling. The damping length due to the synchrotron cooling 
is given by $l_{syn}\sim3\cdot10^{4}(B/10^7\mathrm{G})^{-2}(\Gamma_{\perp}/10^2)^{-1}
(\sin\theta_p/0.06)$~cm, where $\Gamma_{\perp}=\Gamma/\Gamma_{||}$
and $\Gamma_{||}=1/\sin\theta_p$. The charge particles quickly lose
their perpendicular momentum  via the
synchrotron radiation. 
The minimum Lorentz factor
 of the pairs in the magnetosphere may be described by 
 $\Gamma\sim \Gamma_{||}\sim 17$, which
is corresponding to the synchrotron characteristic energy  
$E_{c}\sim 3 (\Gamma/17)^2(B/10^7~\mathrm{G})(\sin\theta_p/0.06)$~eV.
 In the present model, therefore,
the spectral index $s_{\nu}$ of the synchrotron photons varies from
$s_{\nu}=(p-1)/2$ with $p=2$  of equation (\ref{dist})
 in X-ray bands to $s_{\nu}\sim-1/3$, which is
reflecting the single particle emissivity,  
 below the energy $E_c$.

\subsubsection{emission direction of the pairs}
\label{emid}
We use the rotating dipole field in the inertial  
observer frame (hereafter IOF). On the other hand, 
the previous works such as CRZ00 and DHR04 used the rotating 
dipole field in the co-rotating frame, in which the emission direction
coincides with the local magnetic field direction,  and  
performed the Lorentz transformation to calculate the emission direction
in IOF. As a result, the configuration of the magnetic field in IOF 
is different between the present model and the previous works,
although the difference is small except for near the light cylinder. 
 
For a high Lorentz factor, we can anticipate
 that the emission direction of the particles  coincides with the
direction of the particle's velocity. In IOF, the motion of the pairs
created outside of the gap may 
be described by 
\begin{equation}
\mbox{\boldmath$n$}=\beta_0\cos\theta_p\mbox{\boldmath$b$}
+\beta_0\sin\theta_p\mbox{\boldmath$b$}_
{\perp}+\beta_{co}\mbox{\boldmath$e$}_{\phi},
\label{pmotion}
\end{equation}
where the first term in the right hand side represents the particle
motion along the field line,  $\mbox{\boldmath$b$} 
=\mbox{\boldmath$B$}/B$ (or $-\mbox{\boldmath$B$}/B$) 
for the particles migrating parallel (or counter parallel) to the direction
of the magnetic field. In this paper, we consider only 
outgoing particles because the photons emitted by ingoing particles
will be much fainter than that by the outgoing particles (CRZ00). 
 The second term in equation (\ref{pmotion}) represents gyration motion 
around the magnetic field line and the third term is co-rotation 
motion with the star,
$\beta_{co}=\rho\Omega/c$. The unit vector $\mbox{\boldmath$b$}_{\perp}$
perpendicular to the magnetic field line becomes 
\begin{equation}
\mbox{\boldmath$b$}_{\perp}\equiv\pm(\cos\delta\phi\mbox{\boldmath$k$}+\sin\delta\phi\mbox{\boldmath$k$}
\times\mbox{\boldmath$b$}),
\label{gyration}
\end{equation}
where $\pm$ represents the gyration of the positrons ($+$) and the
electrons ($-$), $\delta\phi$ refers the phase of gyration motion and 
$\mbox{\boldmath$k$}=(\mbox{\boldmath$b$}\cdot\nabla)\mbox{\boldmath$b$}/
|(\mbox{\boldmath$b$}\cdot\nabla)\mbox{\boldmath$b$}|$ is 
the unit vector of the curvature of the magnetic field lines. 
The gyration phase $\delta\phi$ is defined so that the value 
increases in the 
direction of the gyration motion of the positrons and  
 so that the pairs with $\delta\phi=0$
emit the photons in the plane spanned by the directions of 
 the local magnetic field and  it's curvature if there were no
co-rotation motion in equation (\ref{pmotion}). The co-rotation
motion affects the emission direction as the aberration effect.
 The value of $\beta_0$ at each point is determined by the condition that
$|\mbox{\boldmath$n$}|=1$.

The emission direction of equation (\ref{pmotion}) 
is described in terms of the
viewing angle measured from the rotational axis, $\xi=\cos^{-1}n_z$, 
and the rotation phase, $\Phi=-\Phi_{n}-\mbox{\boldmath$r$}\cdot\mbox{\boldmath$n$}$, where $n_z$ is the component of
the emission direction parallel to the rotational axis, $\Phi_{n}$
is the azimuthal angle of the emission direction and
$\mbox{\boldmath$r$}$ is the emitting location in units of the light radius.

Because the particles distribute on the gyration phase
$\delta\phi$, the emitted beam
at each point must  become cone like shape with opening angle $\theta_p (r)$.
For each radiating point, the emission directions of the 
different particles on the gyration
phase are projected onto the different points in $(\xi,\Phi)$ plane. 
Furthermore, the polarization plane of the radiations also 
depends on the gyration phases of the radiating particles. 
Taking into account the dependence on the gyration phase, therefore, 
we calculate the radiations from the particles for all of the 
 gyration phase $\delta\phi= 2\pi i/n$ ($i=1,\cdots,n-1$).
 Figure~\ref{map} shows the emission projection onto
$(\xi,\Phi)$ plane for different gyration phases $(\delta\phi=0^{\circ},
~90^{\circ},~180^{\circ}$ and $270^{\circ})$ for the positrons. For
the electrons, we find from the equation~(\ref{gyration}) that  
the emission projection map of the electrons is identical with 
Figure~\ref{map} but the gyration phase is different by 
$180^{\circ}$; 
for example, the panels for
$\delta\phi=0^{\circ}$ and $90^{\circ}$ for the positrons in
Figure~\ref{map}
 also describe the projection maps of emission from the electrons with 
$\delta\phi=180^{\circ}$ and $270^{\circ}$, respectively.  

 With the projection map of the emissions, the  expected pulse
profile is determined by choosing the viewing angle $\xi$ of the
observer and collecting all photons from the possible emitting points
and the gyration phases  with the emissivity of equation
(\ref{emis}). 

\subsection{The Stokes parameters}
\label{stokes}
We assume that the radiation at each point  linearly polarizes
with degree of $\Pi_{syn}=(p+1)/(p+7/3)$, where $p$ is the power law index
of the particle distribution, and circular polarization is zero,
that is, $V=0$ in terms of the Stokes parameters. The direction of the 
electric vector of the electro-magnetic wave toward the observer 
is parallel to the projected direction of the acceleration 
of the particle on the sky (Blaskiewicz et al. 1991). 
 The magnitude of the microscopic acceleration of
 the gyration motion is much larger than that of the macroscopic
acceleration of the co-rotation motion so that
the ratio of the magnitude of the two accelerations becomes 
 $\omega_B/\Omega\sim 10^8 (B/10^6\mathrm{G})
(\Gamma/10^3)^{-1}$, where $\omega_B$ and $\Omega$ are the gyration
and the co-rotation frequencies, respectively. Unless the pitch angle is very
small, the acceleration with  equation (\ref{pmotion}) 
is approximately written  by 
\begin{equation}
\mbox{\boldmath$a$}\sim \beta_{0}\omega_B\sin\theta_p(-\sin\delta\phi\mbox{\boldmath$k$}
+\cos\delta\phi\mbox{\boldmath$k$}\times\mbox{\boldmath$b$}).
\label{accele}
\end{equation}
The electric vector $\mbox{\boldmath$E$}_{em}$ 
emitted in the direction $\mbox{\boldmath$n$}$  becomes 
$\mbox{\boldmath$E$}_{em}\propto \mbox{\boldmath$a$}
-(\mbox{\boldmath$n$}\cdot\mbox{\boldmath$a$})\mbox{\boldmath$n$}$.

To calculate the Stokes parameters $Q^i$ and $U^i$ for each radiating 
point,  we define the position angle $\chi^i$ to be angle between 
the electric field $\mbox{\boldmath$E$}_{em}$  and the projected
rotational axis on the sky, 
$\mbox{\boldmath$\Omega$}_p=\mbox{\boldmath$\Omega$}-
(\mbox{\boldmath$n$}\cdot\mbox{\boldmath$\Omega$})\mbox{\boldmath$n$}$. The Stokes parameters $Q^{i}$ and
$U^{i}$ at each radiation is represented by
$Q^{i}=\Pi_{syn}I^{i}\cos2\chi^{i}$ and
$U^{i}=\Pi_{syn}I^{i}\sin2\chi^{i}$, where $I^{i}$ is the
intensity. After collecting the photons from the possible points for
each rotation phase $\Phi$ and  a viewing angle $\xi$, the expected
p.d. and p.a. are, respectively,  obtained from 
\begin{equation}
P(\xi,\Phi)=\Pi_{syn}\frac{\sqrt{Q^2(\xi,\Phi)
+U^2(\xi,\Phi)}}{I(\xi,\Phi)},
\end{equation} 
and 
\begin{equation}
\chi(\xi,\Phi)
=0.5\mathrm{atan}
\left[\frac{U(\xi,\Phi)}{Q(\xi,\Phi)}\right],
\end{equation}
where $Q(\xi,\Phi)=\Sigma Q^i$ and $U(\xi,\Phi)=\Sigma U^i$
are the Stokes parameters after collecting the photons.

Finally, we describe the difference between polarization characteristics
predicted by the curvature emission and synchrotron emission
models. If we ignore the effects of the aberration due to the
co-rotation motion, the direction of the 
electric vector of the wave for the curvature and synchrotron cases are,
respectively, parallel to and  perpendicular to the  magnetic field projected 
 on the sky (Rybicki \& Lightman 1979). Secondly,  
in the curvature radiation model 
the photons are radiated only one direction at each point, which 
coincides with the direction of the local magnetic field line if we 
ignore the aberration effect. Furthermore, the radiations from 
neighboring  positions polarize  in similar directions. In such a case, 
 the intrinsic level of the polarization degree at the each
radiating point remains in  the bridge and off-pulse phase of the
light curve (DHR04). Therefore, to explain the observed polarization
degree $\sim10\%$ at the bridge phase of the Crab pulsar, 
the curvature emission model may require 
the radiation linearly polarized  with about 10\% at each radiating
points. As we have mentioned for the synchrotron
case, on the other hand, the
photons are emitted  along the surface of the cone with opening angle 
$\theta_p$ at each radiating position. Furthermore,  
the polarization plane of the  radiation depends on the gyration phase
of the radiating particle. In such a case, the 
observed radiation  consists of  the radiations from the 
different particles on the gyration phase. This overlap of the
radiations causes a strong  depolarization, and as a result  a lower 
polarization degree is expected for the synchrotron case. We need not 
assume the radiations with a low polarization degree at each
radiating point to explain the Crab data.  In the present model, the
intrinsic polarization degree of the radiation at the each radiating
point is $\sim70\%$ using the particle distribution $p=2$ described by
equation (\ref{dist}).

\subsection{Model parameters}
\label{para}
In this subsection, we introduce the model parameters. 
The  inclination angle $\alpha$
 and the viewing angles $\xi$ measured from the rotational axis 
are the  model  parameters. Because the inner boundary of the outer gap
is determined by the current through the gap (Takata et al. 2004), 
we consider the position of inner boundary located inside of
 the null charge. In this paper, 
 the ratio of the radial distances to the
inner boundary $r_i$ and the null charge surface of the rotating dipole
$r_n$, which is a function of the azimuthal angle, 
 is teated as a model parameters, and is assumed to be constant for
each field line as described in section~\ref{oustruc}.

Since the magnetic field must be modified by
the rotational and the  plasma effects near the light cylinder,  
the last-open field line
must be different with the traditional magnetic surface 
that is tangent to the light cylinder for the vacuum case. For
example, Romani (1996) defined the last open lines as the field lines
parallel to the rotational axis at $r=R_{lc}/2^{1/2}$. where the corotational
velocity equals the Alfv\'{e}n speed. 
 To specify the gap upper surface, therefore,  
it is convenient to refer the footpoints of the magnetic field lines on 
the stellar surface. With the
assumption that the gap upper surface coincides with a magnetic
surface, we parameterize the fractional polar angle 
$a=\theta_u/\theta_{lc}$,
 where $\theta_u$ is the polar angle of the footpoints of the  magnetic
field lines of the gap upper surface and $\theta_{lc}$ is the polar angle
of the field lines which are tangent to the light cylinder for the
vacuum case.  

Finally, we describe how the model parameters ($\alpha$, $\xi$, $r_i$
and $a$) are diagnosed by the
present model and the Crab data. The model parameters are chosen so
that the expected light curve, the spectrum and the polarization
characteristics are simultaneously consistent with the Crab data such
as the phase separation $\delta\Phi\sim0.4$~phase between two peaks and
the large position angle swings at the both peaks. 
As we will demonstrate in section~\ref{lcurve},  the features of the expected 
light curve is  sensitive to the viewing angle $\xi$ 
but not to the position of the inner boundary $r_i$,  
if we fix the inclination angle $\alpha$ and the fractional angle
$a$.  Therefore, the viewing angle $\xi$ is determined by comparing
the model and  the observed light curves. 
 With the determined viewing angle, on the other
hand, the radial distance to the inner boundary  $r_i$ affects
sensitively to  the polarization characteristics at the off-pulse phase. 
Therefore, if the inclination angle $\alpha$ and the gap upper surface
$a$ are determined in some way, the viewing angle $\xi$ of the observer
and the position of the inner boundary of the outer gap $r_i$
 are diagnosed by the present model and the Crab data. 
The present geometrical model  produces a consistent spectrum with the
Crab data (section~\ref{polari}) using  the viewing angle determined from
the observed light curve.
We need not introduce another model parameter for fitting the
spectrum. It is difficult to constrain  
both the inclination angle $\alpha$ and the upper surface $a$ 
with the present model.  As we will show in section~\ref{inmag},  however, 
if  either inclination angle $\alpha$ or the altitude of the upper surface 
 $a$ is determined in some way, the other may be diagnosed  
by the present model. 

\section{RESULTS}
\label{result}
\subsection{Light curve}
\label{lcurve}
\begin{figure}
\begin{center}
\includegraphics[width=15cm, height=15cm]{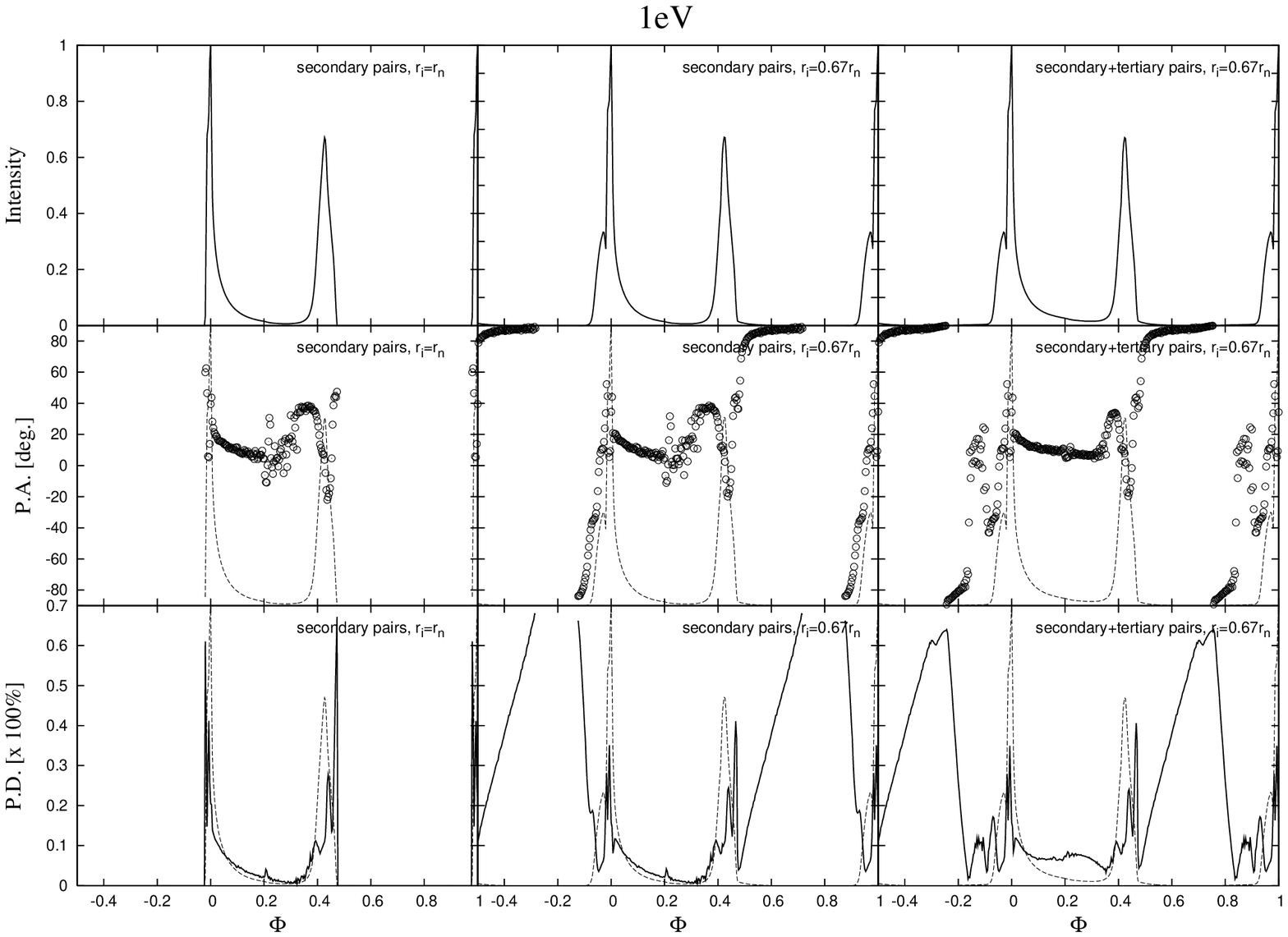}
\caption{Polarization characteristics for three different emission
geometries. The left column shows the result for the traditional model, which
considers the radiation from the
secondary pairs and emission regions extending from the null charge
surface, $r_i=r_n$. The middle and right column take into account the
radiations from the inside the null charge surface, $r_i=0.67r_n$, 
and furthermore the right column considers the effects of 
 the emissions from the tertiary pairs. The upper, middle and lower
panels in each column show, respectively, the light curve, the position
angle and the polarization degree. The model parameters are
$\alpha=50^{\circ}$, $\xi=100^{\circ}$ and $a=0.94$.}
\label{compari}
\end{center}
\end{figure}

Figure~\ref{compari} compares the polarization characteristics at
1~eV predicted by three different emission geometries. The left column
summarizes the results for the traditional outer gap geometry, in
which the inner boundary of the gap is located at the 
null charge surface of the Goldreich-Julian charge density. 
Although the traditional model in
CRZ00 and the present model assume the different extensions of the
outer gap  in the azimuthal direction, that is  around half (in CRZ00)
and whole (in the present model) polar-cap region, we find that 
the predicted polarization
characteristics  are not so different between the two azimuthal
extension of the outer gap
geometries as long as the inner boundary is located at the null 
charge surface.  In this section, therefore, the gap geometry 
that stars from the null
charge surface and extends around whole polar cap region is also
called as "traditional geometry". 

The middle and right in Figure~\ref{compari} columns show the results 
for the radial distance of 67\% of the distance to null charge
 surface $r_i=0.67r$, and  
the right column is taking into account also the emissions
from the tertiary pairs. The other model parameters are
$\alpha=50^{\circ}$, $a=0.94$,
and $\xi\sim100^{\circ}$,  where the viewing angle is
chosen so that the predicted phase separation between the two peaks is
consistent with the observed value $\delta\Phi\sim0.4$~phase.  In the
figure, we define zero of the rotation phase at the main peak.

By comparing the pulse profiles between the light curves of the 
left and middle columns, we find
that the radiations from the secondary pairs  inside the null
charge surface contribute to the outer-wing and the off-pulse 
emissions. In the present case, the off-pulse emissions
are $<10^{-1}\%$ of the peak flux, because the line of sight
marginally passes through the emission regions in the off-pulse phase
as the horizontal lines show in Figure~\ref{map}. Near the inner
boundary, because the accelerating electric field and resultant the
energy of emitted  primary photons in the gap are small,  
the secondary pairs are produced with a lower energy,  and emit the
synchrotron photons with  a smaller emissivity. 

As Figure~\ref{map} indicates, the emerging radiations in the light
curve originate from the two poles; one pole contributes to the light
curve with the two peaks and the bridge photons emitted beyond  
the null charge surface, and the other contributes 
 with the outer-wing and the off-pulse photons emitted 
inside the null charge surface. 
Dyks \& Rudak (2003) have proposed the radiations associated with two
magnetic pole. In that model,  with the constant emissivity 
along the magnetic field lines, the two peaks are associated with 
the different poles and  the different emission regions, and are formed
by the caustic effect near the stellar surface. 
In the present outer gap model, on the other hand, the
electric field, and the resultant  emissivity of the synchrotron
radiation of the secondary pairs quickly decreases inside 
the null charge surface. Therefore, 
although the caustic effect near the stellar surface is  strong, 
the emissions inside  the null charge surface do not make a strong
peak compared  with the present main peak, which is
 formed by the radiations near the light cylinder. 
 In the present case, therefore,
the two peaks in the light curve is 
associated with the one magnetic pole. 

By comparing the flux levels of the
bridge emissions between the light curves in the middle and right columns, we
find that the tertiary pairs mainly contribute to the
emissions at the bridge phase. 
This is because the tertiary pairs are born and emit photons at higher
altitude than the secondary pairs, which make two peaks in the light curve.

\subsection{Polarization}
\label{polari}
 Middle and lower panels in Figure~\ref{compari} 
show the predicted polarization position angle (p.a.) 
and the polarization degree (p.d.), respectively. 
For reference, the light curve is overplotted in each frame.  

\begin{figure}
\begin{center}
\includegraphics[width=15cm, height=15cm]{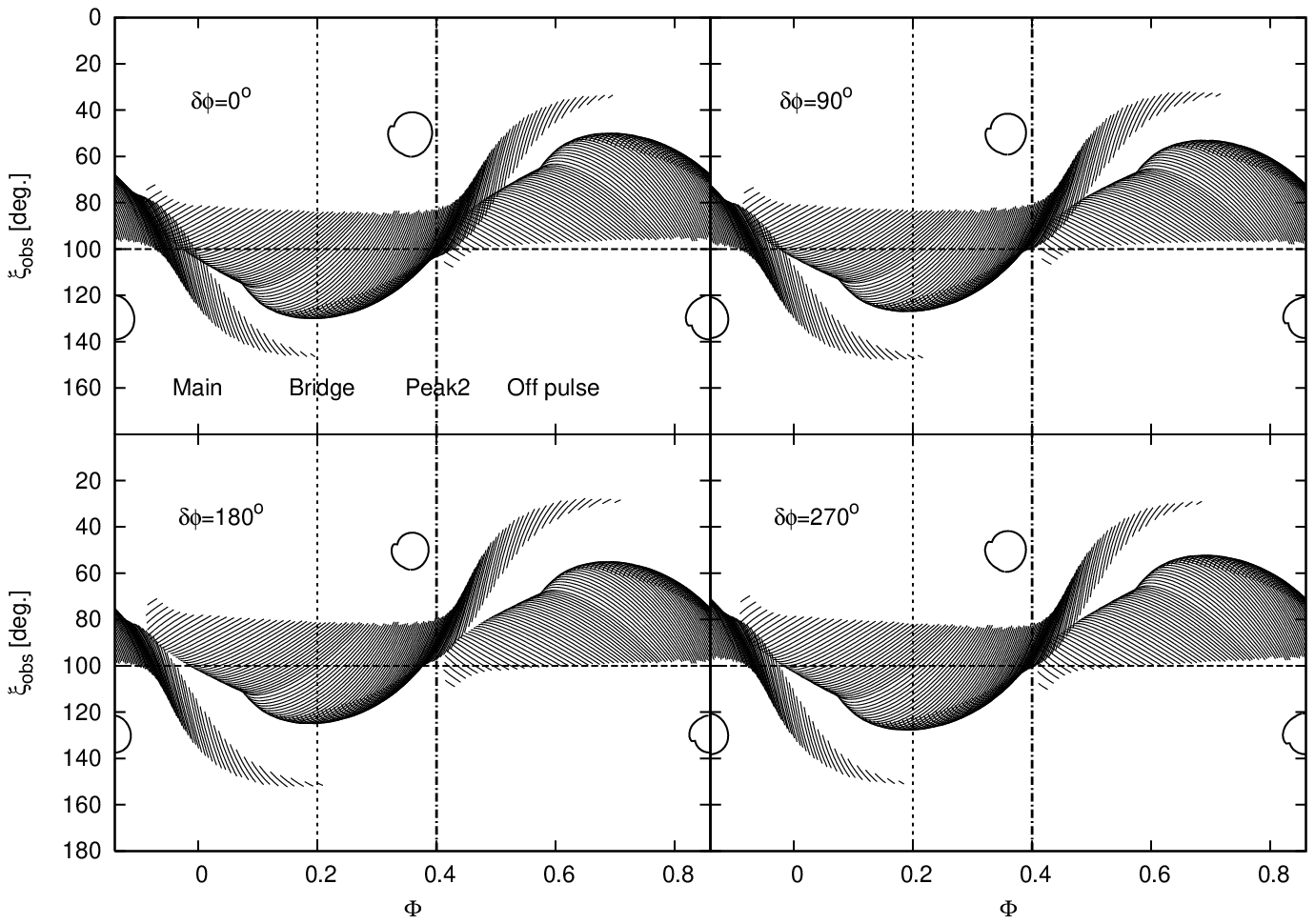}
\caption{Emission projection onto ($\xi,~\Phi)$ plane for the
magnetic surface, $a=0.94$, of the upper boundary of the outer gap. 
The  inclination angle is $\alpha=50^{\circ}$. 
The emission region extends from $r=0.67r_n$ to
$r=R_{lc}$ or $\rho=0.9R_{lc}$. 
Each panel shows the radiations from the positrons with
the gyration phase of $\delta\phi=0^{\circ},~90^{\circ},~180^{\circ}$
and $270^{\circ}$, respectively. The thick solid circles 
in the figure show the shape of the polar cap. }
\label{map}
\end{center}
\end{figure}

As  seen in the polarization characteristics by the traditional model,
we find that the secondary emissions beyond the
null charge surface  make the polarization characteristics  such  
that the polarization degree takes a lower value 
at the bridge phase and a larger value near the peaks. In the
synchrotron case, the cone like beam is radiated at each point, and an
overlap of the radiations from the different  particles on the  gyration
phase causes the depolarization. For the viewing angle
$\xi\sim100^{\circ}$,  the radiations from all gyration phases contribute to the observed
radiation at the bridge phase as the vertical dotted lines 
at $\Phi=0.2$~phase in Figure~\ref{map} show. 
In such a case, the depolarization is strong, and as a result, the
emerging radiation from the secondary  pairs polarizes with  a very low p.d. 
($<10\%$).  Near the
peaks, on the other hand, the radiations from the some gyration 
phase are not observed as the vertical dotted-dashed 
lines in Figure~\ref{map} show. For example, the observer with the
viewing angle $\xi\sim100^{\circ}$ detects the photons from the
gyration phase $\delta\phi=0^{\circ}$ at the rotation phase
$\Phi=0.4$~phase (second peak), but does not detect from 
 $\delta\phi=180^{\circ}$. In such a case, the depolarization is
weaker and the emerging  radiation  highly polarizes.

As the polarization degrees in the middle and right columns in 
Figure~\ref{compari} show,
 the radiations from the inside the null charge surface  may be 
 observed with a large polarization degree at the off-pulse phase. 
This is because the line of sight  $\xi\sim 100^{\circ}$ passes
through  marginally the edge of
the radiating region with $r_i=0.67r_n$  at the off-pulse phase
as horizontal lines in  Figure~\ref{map} show. 
The observer can not detect the radiations 
 from the particles within a range of the gyration phase; for example, at the
rotational phase $\Phi=0.6$~phase (off-pulse phase) in
Figure~\ref{map},  the observer with viewing angle
$\xi\sim100^{\circ}$ detects the radiations from the particles with
the gyration phases $\delta\phi=180^{\circ}$ and
$\delta\phi=270^{\circ}$, but does not with $\delta\phi=0^{\circ}$ and
$\delta\phi=90^{\circ}$. In such a case, the depolarization 
in the off-pulse phase  is weak and the expected p.d. exhibits 
a larger value.  Actually, as we will show in section~\ref{inner}, the
p.d. at the off-pulse phase is sensitive to the position of the inner
boundary $r_i$.

We can see the effects of the tertiary pairs on the polarization
characteristics at the bridge phase.  
By comparing the p.d. between middle and right panels,
 we find that tertiary pairs produce the radiations with 
 $\sim 10\%$ of the p.d. at the bridge phase. 

\subsection{Comparison with observations}
\label{comob}
\begin{figure}
\begin{center}
\includegraphics[width=15cm, height=15cm]{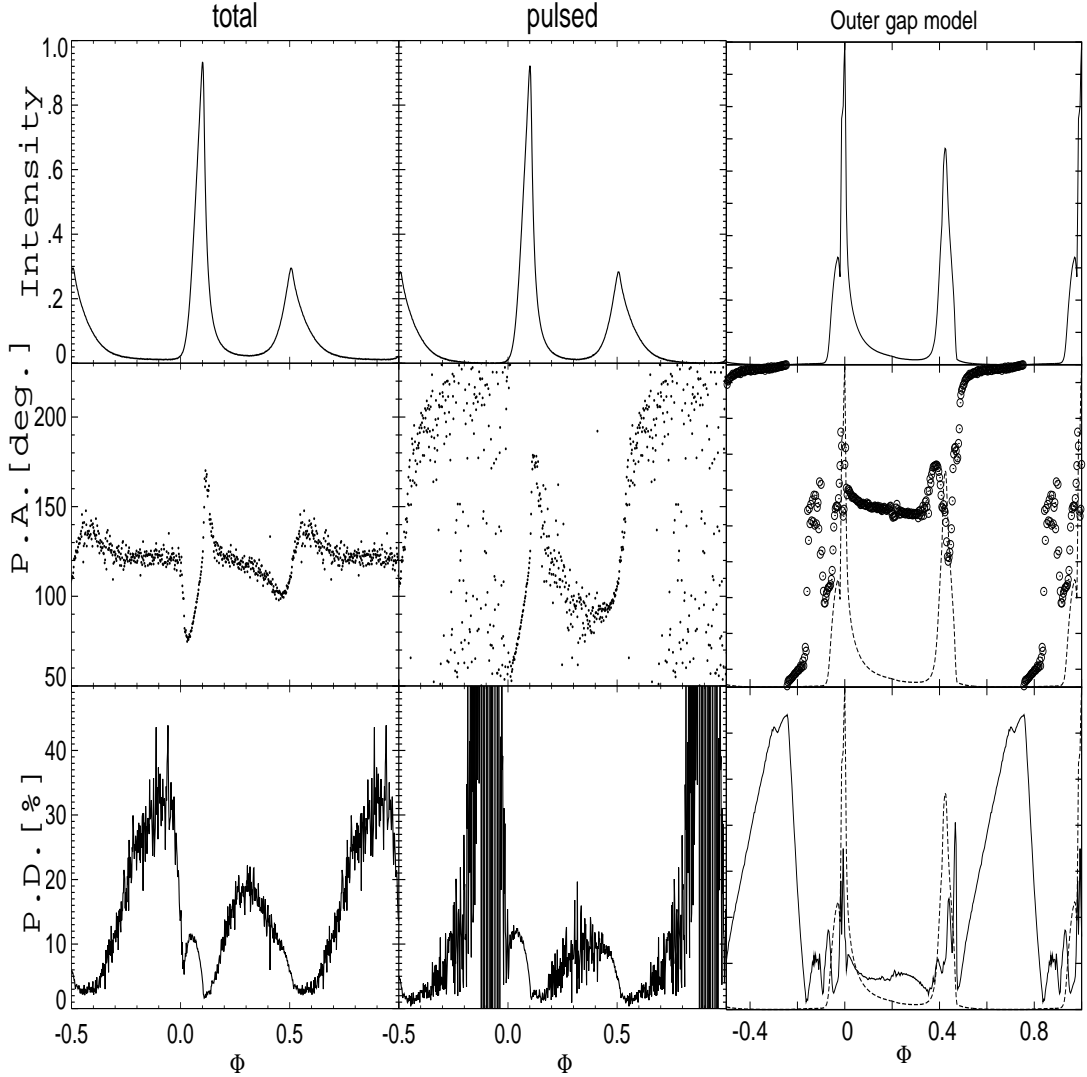}
\caption{Optical polarization for the Crab pulsar. Left:Polarization 
characteristics for the total emissions from the Crab
pulsar. Middle:Polarization characteristics of the emissions after
subtraction of the DC level (Kanbach et al. 2005). Right:Predicted polarization
characteristics at 1~eV for $\alpha=50^{\circ}$, $a=0.94$, $\xi\sim100^{\circ}$
and $r_{i}=0.67r_n$. The figures for the Crab optical data 
was transcribed from  DHR04 and was arranged by authors.}
\label{datmod}
\end{center}
\end{figure}

Figure~\ref{datmod} compares the predicted  polarization 
characteristics at 1~eV with the Crab optical data. 
Left and middle columns are, respectively,  the Crab optical data for the total
emissions and for the emissions after subtraction of the DC level, 
which has the constant intensity at the
level of 1.24\% of the main pulse intensity (Kanbach et al. 2005).

In the total emissions (left column), the impressive
polarization feature from the Crab pulsar is that the off-pulse and bridge
phases have the  fixed value of the p.a. These polarization features of the
observation are not predicted by the present model, 
which predicts about $90^{\circ}$ difference on the p.a. between
 the off-pulse and the bridge phases as the right column  shows.

The constant p.a. in the total emissions may suggest that 
the Crab optical emissions 
consist of two components, that is,  constant and pulsed
components. The DC level emissions may include both the magnetospheric
component and the background components (e.g. the pulsar wind and the
nebula components). After the subtraction of the DC level, 
the large p.a. swings larger than 
$\sim100^{\circ}$ appears at the both peaks, 
and the constancy of the p.a. at both bridge and off-pulse phase
disappears.  We can see that the polarization characteristics predicted
by the present model are more consistent with the Crab optical data 
after the subtraction of the DC level.  
Especially, the model reproduces the most striking feature in
 the observed p.a. that the large swing at both peaks, and 
the  observed low p.d. at bridge phase $\sim 10\%$. Also, 
the pattern of the p.d. are reproduced by the present model. 

In the off-pulse phase, it is not clarified that which radiation
component, that is , the magnetospheric or back ground  
(e.g the pulsar wind and nebula components) components dominates
the other one at the off-pulse phase, where the flux level
of the magnetospheric component is much 
 smaller than the peak flux. 
 Furthermore, the data for emissions after
subtraction of the  DC level has a few photons  in the  off-pulse
phase so that  the polarization behavior in the off-pulse phase may
not be determined. The present model (e.g. the polarization characteristics
in the right column in Figure~\ref{compari}) predicts that radiation 
from the magnetospheric component  at the off-pulse phase has a relatively
constant position angle, which is  about $90^{\circ}$ difference 
from  that in the bridge phase. 
This constancy of the p.a. in the off-pulse phase
  is  not sensitive to the model parameters 
such as the viewing angle $\xi$ and the 
position of the inner boundary $r_i$  
(Figures~\ref{compari1} and $\ref{compari2}$). 
 
 Although the present model has successfully explained the main features
of the Crab optical polarization data, it also has some 
disagreements with  the Crab  data. 
For example,  the model predicts the small p.a. swing  at the
leading-wing of the second peak before appearing the large p.a. swing
at the second peak. In the p.d., furthermore, the another peaks in the
p.d.  at both peaks are predicted.

\begin{figure}
\begin{center}
\includegraphics[width=10cm, height=10cm]{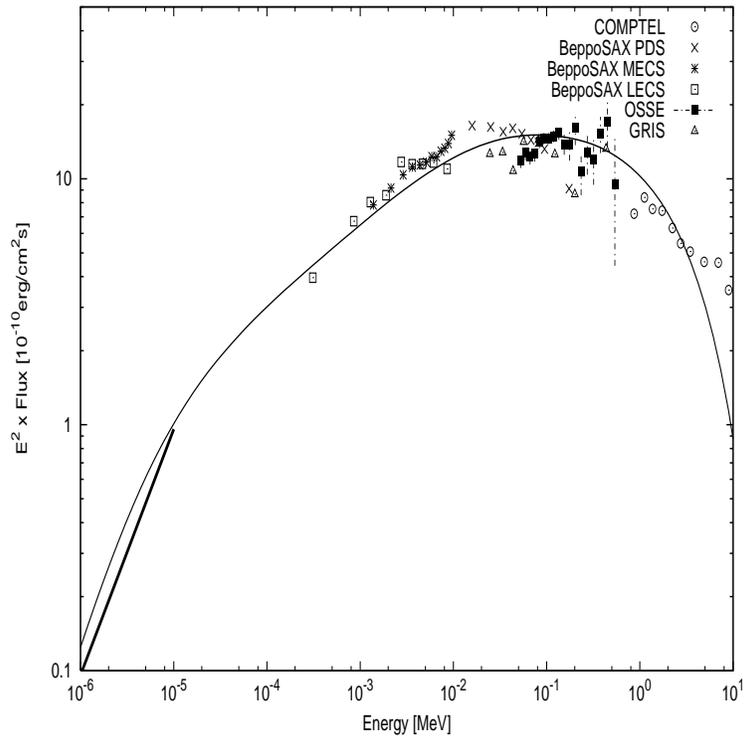}
\caption{The optical-X ray spectrum for the Crab pulsar. Thin solid
line shows the expected spectrum for $\alpha=50^{\circ}$,
$a=0.94$, $r_i=0.67$ and $\xi_i\sim100^{\circ}$. The X-ray data  are taken
from Kuiper et al. (2002) and reference therein,  
and the optical data from Sollerman et al. (2000).}
\label{spectrum}
\end{center}
\end{figure}

Figure~\ref{spectrum} compares the model spectrum with the Crab data 
in optical-MeV bands.  The model parameters are same with that in the
right column in Figure~\ref{datmod}. In this case. we assume 
that the pairs escape from the light
cylinder with the Lorentz factor $\Gamma\sim17$
(section~\ref{cooling}), which predicts the spectral break around 10~eV. 
The model spectrum also explains the general features of
the data. Therefore, the outer gap model can explain the general
features of the observed light curve, the spectrum and the
polarization characteristics in optical band for the Crab pulsar,
simultaneously.

\subsection{Dependence on the model parameters}
\label{depend}
In this section, we discuss the dependence of the polarization
characteristics on  the model parameters  and diagnose that for the
Crab pulsar. 

\subsubsection{viewing angle}
\label{viewing}
\begin{figure}
\begin{center}
\includegraphics[width=15cm, height=15cm]{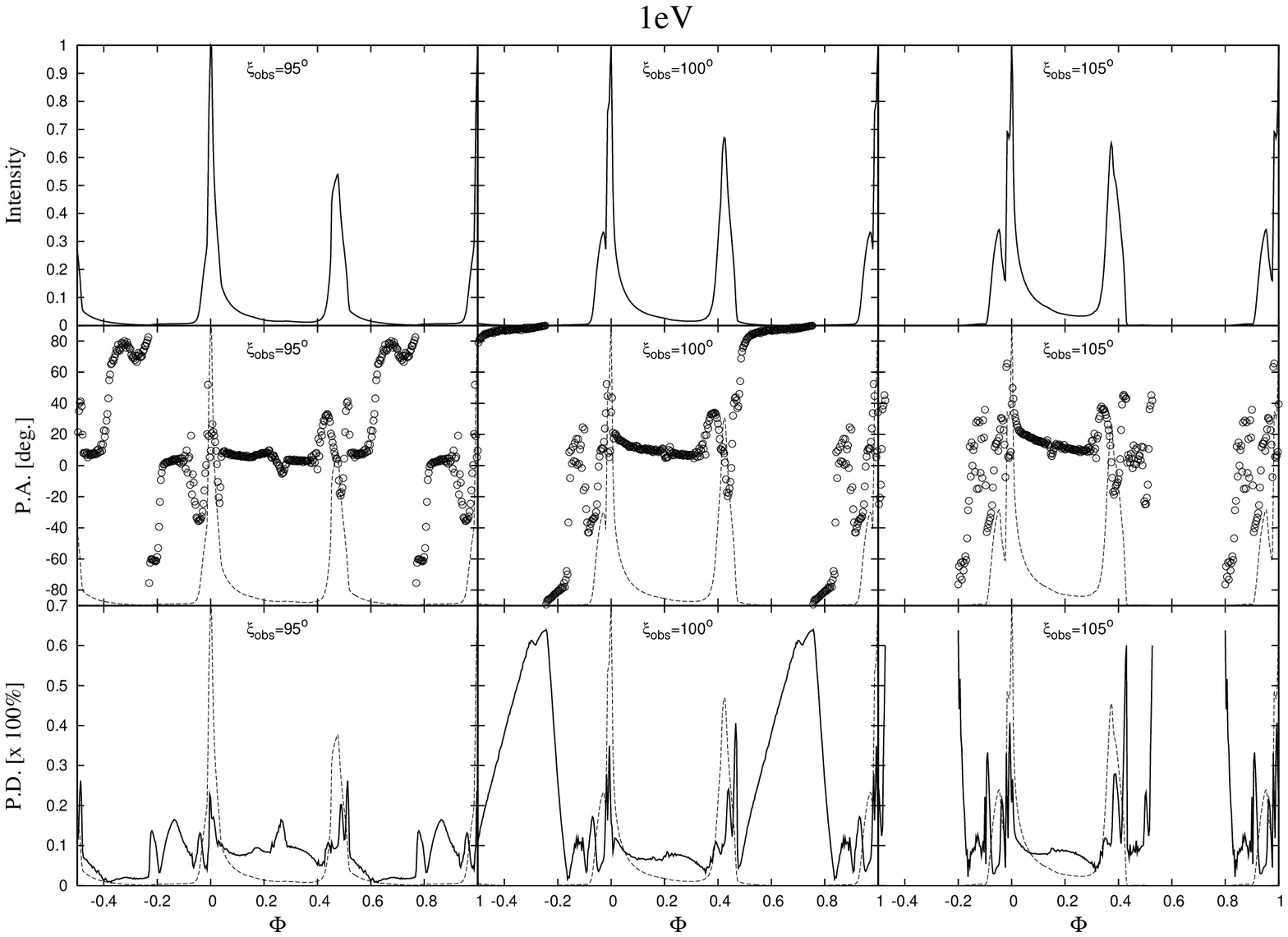}
\caption{The polarization characteristics for three different viewing
angles, $\xi\sim95^{\circ}$ (left column), $100^{\circ}$ (middle column)
and $105^{\circ}$ (right column), for
$\alpha=50^{\circ}$, $a=0.94$ and $r_i=0.67r_n$.}
\label{compari1}
\end{center}
\end{figure}

Figure~\ref{compari1} summarizes the dependence of the polarization
characteristics on the viewing angle. With $\alpha=50^{\circ}$, 
$a=0.94$ and $r_{i}=0.67r_n$, the left, middle and right
columns show, respectively,  the polarization characteristics  
 for the viewing angle of 
$\xi\sim95^{\circ}$, $\xi\sim100^{\circ}$ and
$\xi\sim105^{\circ}$. For $\xi\sim100^{\circ}$, the phase separation of
the two peaks in the light curve is 
$\delta\Phi\sim0.4$~phase similar with the observation.

We can see  that the phase separation becomes wider (or narrower) with 
decreasing (of increasing ) the viewing angle from 
$\xi\sim100^{\circ}$. For example,
as the light curve of the left column shows, the phase separation
 between two peaks for $\xi\sim 95^{\circ}$ 
becomes obviously wider than $\delta\Phi\sim0.4$~phase. 
On the other hand, the phase separation 
for $\xi\sim 105^{\circ}$ becomes narrower than the data.
 Furthermore, in the light curve for
$\xi\sim105^{\circ}$, we can see  a conspicuous peak in the leading-wing of
 the main  peak. This leading small peak is formed by the radiations inside the
null charge surface of the radiation region connecting to the other pole,
while the main peak is formed by the radiations  near the light cylinder. 
For $\xi\sim 95^{\circ}$ and $\sim100^{\circ}$,
 these  two peaks are observed as a single main peak,  because the phase
separation of the two peaks is very narrow. 

By comparing the p.d. for three cases, we find that 
p.d. in off-pulse phase increases with the viewing
angle; e.g. typical p.d. in the off-pulse phase is $\sim20\%$ for
$\xi\sim95^{\circ}$ and $\sim60\%$ for $\xi\sim100^{\circ}$.
 As we have mentioned in section~\ref{polari}, 
the present model predicts highly polarized radiations 
at the off-pulse phase, if the line of sight marginally  
passes through  the emission region. Increasing the viewing angle with
a specific  position of the inner
boundary $r_i$, the line of sight approaches
the inner boundary from inside the emission region as we can expect
from  Figure~\ref{map}, and then 
the radiations from wider range of the gyration
phases become to be beamed out of the line of the sight. Therefore, 
the p.d. in off-pulse phase increases with the viewing angle. 
Finally, if the line of sight passes through outside the
emission region, there are no emissions in the off-pulse phase such as
the light curve of the right column in Figure~\ref{compari1}. 
  
As we have seen, the viewing angle affects sensitively to the model
light curve and the polarization degree in the off-pulse
phase. In the present model, especially, the observed phase separation of the two peaks
$\delta\Phi\sim0.4$ restricts the viewing angle with 
$\pm5^{\circ}$ uncertainty.

\begin{figure}
\begin{center}
\includegraphics[width=15cm, height=7cm]{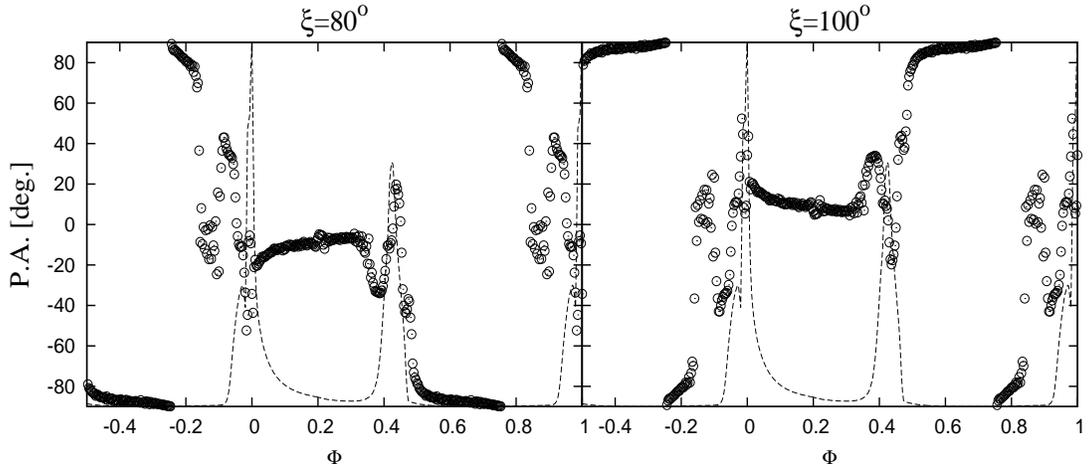}
\caption{The polarization position angle for viewing angles,
$\xi\sim80^{\circ}$ and $100^{\circ}$,  which are mutually symmetric
with respect to the rotational equator. The calculations are for
$\alpha=50^{\circ}$, $a=0.94$ and $r_i=0.67r_n$. For the reference,
the light curve are overplotted.}
\label{paangle}
\end{center}
\end{figure}

Let us consider the two viewing angles mutually symmetric with respect to 
the rotational equator (e.g. the viewing angles $80^{\circ}$ and
$100^{\circ}$). For such symmetric viewing angles,   
the  light curves, the spectra and the p.d. curves 
are identical.  However, the p.a. curves 
are mirror symmetry with respect to the rotational
equator because  of the difference directions of 
the projected magnetic field on the sky. 
Figure~\ref{paangle} shows the polarization position angles for the 
viewing angles $\xi\sim80^{\circ}$ (left column) and $\xi\sim100^{\circ}$
(right column) with $\alpha=50^{\circ}$, $a=0.94$ and $r_i=0.67r_n$. 
  By comparing the swing pattern at the both
peaks between the model results in Figure~\ref{paangle} 
and the data in Figure~\ref{datmod}, we find that the p.a. for $\xi\sim100^{\circ}$ is more consistent  with the
Crab data. The behavior of the p.a. swing  does not change for  
different viewing angles  in the same hemisphere. 
 Therefore, the viewing angle  larger than $90^{\circ}$ 
are preferred for the Crab pulsar.

Although we can distinguish the two viewing angle mutually symmetric
with respect to the rotational equator, the present model does not allow to
distinguish the two inclination angles mutually symmetric
with respect to the rotational equator, that is, $\alpha$ and
$180^{\circ}-\alpha$. This is because the present model has
considered the only outgoing electron and positron pairs.
 We have not used the information of 
the magnetic polarity for the emissions from the pairs.

\subsubsection{inner boundary, $r_i$}
\label{inner}
We consider the dependence of the polarization characteristics on
the position of the 
inner boundary of the outer gap by fixing the inclination angle,
$\alpha=50^{\circ}$, the gap upper surface $a=0.94$
and the viewing angle $\xi\sim100^{\circ}$. As mentioned in 
section~\ref{oustruc}, we assume the ratio of the radial distances 
to the inner boundary and
the null charge surface, $r_i(\phi)/r_n(\phi)$, is not a function of
the azimuthal angle $\phi$, although $r_n(\phi)$ and $r_i(\phi)$
depend on the azimuthal angle. 

\begin{figure}
\begin{center}
\includegraphics[width=15cm, height=15cm]{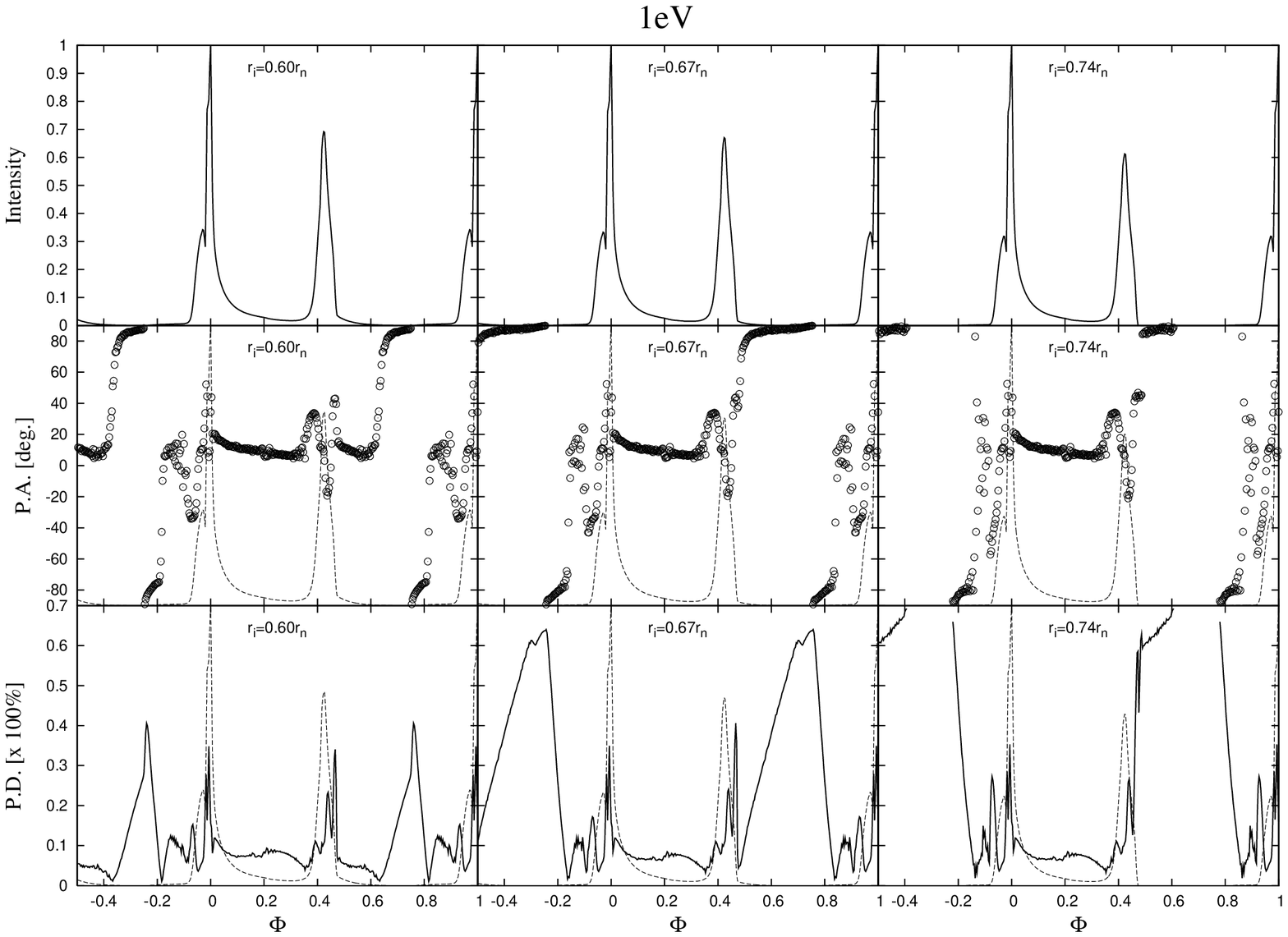}
\caption{The polarization characteristics for three different positions
of the inner boundary, $r_{i}=0.60r_n$ (left column), $0.67r_n$
(middle column) and $0.74r_n$ (right column), for
$\alpha=50^{\circ}$, $a=0.94$ and $\xi\sim100^{\circ}$.}
\label{compari2}
\end{center}
\end{figure}

Figure~\ref{compari2} summarizes the polarization characteristics for 
$r_i=0.60r_n$ (right column), $0.67r_n$ (middle column) and $0.74r_n$
(right column). From  Figure~\ref{compari2}, we find 
that the position of the inner
boundary hardly  affects the expected light curve. 
This is because the radiations from inside
of the null charge surface contribute to the off-pulse emissions
 with a small flux (Figure~\ref{compari}).

From  Figure~\ref{compari2},  we find that the p.d. 
in the off-pulse phase increases with shifting the
inner boundary from the stellar surface toward the null charge 
surface; e.g. typical p.d. in
the off-pulse phase is $\sim30\%$ for $r_i=0.60r_{n}$ and  $\sim60\%$ for
$r_i=0.67r_{n}$.  The reason of the increase is the same with
the results in section~\ref{viewing}, where we discussed the dependence of the
viewing angle $\xi$  with a fixed 
position of the inner boundary $r_i$.  In the present case, the 
line of sight $\xi\sim100^{\circ}$ passes through the radiating points of 
 $r\sim 0.7r_n$ in the  off-pulse phase. In such a case,   for
$r_i=0.6r_n$ the observer
detects the photons from most of all gyration phase, and emerging
radiation polarizes with a low p.d. For $r_i=0.67r_n$, on the other
hand, the line of sight passes through
near the inner boundary  and therefore, 
the radiation at off-pulse phase appears with a large p.d 
as discussed in~\ref{viewing}.  For $r_i=0.74r_n$ (right column),
the line of the sight passes through outside the emission region 
at the off-pulse phase. 

In the middle panels for each column, we see that  the main features of the
p.a. (e.g. the large swings at both peaks) do not depend the position 
of the inner boundary. Therefore, the dependence of the position of the
inner boundary mainly appears as the difference of the p.d. at the
off-pulse phase. In the observation, however, both the emissions from the
magnetosphere and the back ground (e.g. the wind region, and probably
nebula) components would contribute to the off-pulse emissions. 
 It is not clear which component dominates the emissions 
in the off-pulse phase, while the  magnetospheric component must
dominate in the pulse  and the bridge phases. 
To constrain the inner boundary of the emission region with the present
magnetospheric radiation model, the model requires the data of the p.d. of 
the off-pulse emissions from the magnetosphere. If the p.d. of the
magnetospheric component are measured  at the off-pulse phase,  the
present model will  be able to restrict the radial distance
 of the inner boundary better. Furthermore, we may be able to diagnose how
large current runs through the gap with the present geometrical model
and the Crab optical data, because the position of the inner boundary of
the gap is related to the current through the gap (Takata et al. 2004).

As we have seen in sections~\ref{viewing} and \ref{inner}, 
 the polarization characteristics are sensitive to the both 
 viewing angle $\xi$  and the position of the inner boundary
$r_i$, on the other hand, the expected light curve is sensitive to  only
the viewing angle $\xi$. Therefore, the viewing angle is restricted by the
observed light curve rather than the polarization characteristics. With
the viewing angle determined by the light curve, the position of the
inner boundary is restricted by the polarization characteristics.

\subsubsection{inclination angle and gap upper surface}
\label{inmag}
As we have shown in sections~\ref{viewing} and \ref{inner}, 
if the inclination angle
 and the altitude  of the gap upper surface  
could be determined, the viewing angle 
 $\xi$ are restricted by the observed light curve, and  then
 the position of the inner boundary of the gap 
$r_i$ are determined by the Crab optical polarization data. 
In this section, we examine how  $\xi$ and $r_i$ for explaining
the Crab data  are changing with  
the inclination angle $\alpha$ and the altitude of the  gap upper
surface (in other words the fractional angle $a$), above
which the secondary pairs are produced and the emit the
observed photons.
 For each inclination angle $\alpha$ and the altitude of the gap upper
surface $a$,  the viewing angle $\xi$ is chosen to explain the 
observed characteristics
 that the light curve  has two peaks, first peak is stronger than
the second peak (in optical band)  and the phase separation between
two peaks is $\sim0.4$~phase. 
The position of the inner boundary $r_i$ is determined so that 
the p.a. has a large swings at the both peaks, and  the p.d. in the off-pulse
phase becomes about $\sim60\%$ because we do not have
accurate data of the p.d. for the magnetospheric radiations at the
off-pulse phase. 

Table~1 summarizes the expected viewing angle $\xi$ and the
radial distance to the inner boundary $r_i$ for the inclination angles
$\alpha=40^{\circ}$, $50^{\circ}$ and $60^{\circ}$ and the various
 altitude of the gap upper surface. In the table, the increasing of the
value of the fractional angle $a$ means the decreasing of the altitude of
the upper surface of the gap with $f(R_{lc})=0.11$, and of 
the emissions region of the secondary pairs. 
  We find that there is a critical altitude of the upper surface 
($\equiv a_c$)  for each inclination
angle; e.g $a_c\sim0.91$ for $\alpha=40^{\circ}$, 
 $a_c\sim0.93$ for $\alpha=50^{\circ}$ and $a_c\sim0.95$ for
$\alpha=60^{\circ}$. With a fractional angle $a$ smaller  
than the critical value (in other words, with a higher upper surface
than the critical altitude), there are
no viewing angle that produces the light curve consistent with data,
and the expected light curve has triple peaks (leading peak, main peak
and second peak)  
such like the light curve in the right column of Figure~\ref{compari1}.

For a specific inclination angle $\alpha$, the expected viewing angle
$\xi$ increases with decreasing the altitude of the gap upper surface
(or with  the increasing the fractional angle $a$). The reason is 
explained as follows. Firstly, the phase separation of
the two peaks increases with decreasing emission height 
 because the area of the magnetic surface
for the radiation regions becomes wider for lower altitude; for
example,  the viewing angle $\xi\sim100^{\circ}$ produces 
 the phase separation $\delta\Phi\sim 0.4$~phase in the light curve
 with $\alpha=40^{\circ}$ and $a=0.91$, and therefore predicts 
a wider phase separation of the two peaks  than
$\delta\Phi=0.4$~phase for   $a=0.92$. Secondarily, the phase separation 
of the two peaks becomes narrower 
with increasing the viewing angle as shown in Figure~\ref{compari1}. 
As a result, the suitable viewing angle $\xi\sim102.5^{\circ}$
 of $a=0.92$ is larger than $\xi\sim100^{\circ}$ 
of $a=0.91$.

We also find that the critical altitude of the upper surface  
decreases when the inclination angle $\alpha$ increases.  
As we have discussed in section~\ref{lcurve}, if
the phase separation between the main peak and the small leading peak,
which originates of the emissions inside null charge surface, is
enough narrow,  these two peaks appear as a single main peak in the
light curve. If it is not, the light curve has a small peak in 
the leading-wing of the main peak. 
With a fixed altitude of the  upper surface $a$, 
greater inclination angle $\alpha$ produces
a wider phase separation between the main peak and the leading 
small peak. On the other hand, a lower altitude of the emission region
with a fixed inclination angle produces a narrower phase separation 
of the two peaks. As a result,  to have a single main peak 
without the leading small peak, 
 a lower altitude of the emission regions of the secondary pairs, 
in other words, a larger value of the fractional angle $a$ of the gap
upper surface is required for a greater inclination angle $\alpha$.

The position of the inner boundary $r_i$, which is chosen so that the p.d. at
the off-pulse phase becomes $\sim 60\%$, 
 approaches to the  stellar surface  with decreasing (or increasing) of 
the altitude of the gap upper surface (or the fractional angle $a$) . As we
discussed in the third paragraph of this section, 
a larger viewing angle $\xi$ is preferred for explaining the phase
separation of the two peaks for a lower altitude of the gap upper
surface. And, the larger viewing angle (for $\xi>90^{\circ}$) detects the
photons emitted from the positions nearer the stellar surface.  
Therefore, the inner boundary is shifted toward stellar surface 
to hold a constant p.d. at the off-pulse phase 
with decreasing (or increasing) of the altitude of the upper surface 
(or the fractional angle $a$).

Finally, the present local model  cannot identify 
the inclination angle $\alpha$ and the altitude of the upper gap
surface, where may be determined  by the observation
or other ways.  However if either the inclination angle or the
gap upper surface is determined, the present model can constrain 
the other one; for example, if
the inclination angle of $\alpha\sim50^{\circ}$ were determined, 
the Crab pulsar should have the gap upper surface, which is located at
lower altitude than that referred by $a\sim0.93$. 
If the inclination angle $\alpha$ and the altitude of the upper
surface are determined, the viewing angle $\xi$ 
and the position of the inner boundary $r_i$
are determined by the observed light curve and the polarization characteristics
as we discussed in sections~\ref{viewing} and ~\ref{inner}.

\section{SUMMARY \& DISCUSSION} 
\label{discuss}

We have considered the light curve and  the spectrum for the Crab
pulsar predicted by the outer gap model which takes into account 
the emissions from the inside the null charge surface and from the tertiary
pairs.  We  have also calculated the polarization characteristics 
in the optical band. We have shown that the emissions from the inside 
the null charge surface contribute to the outer-wing and the 
off-pulse phase. On the other hand, 
the radiations from the tertiary pairs contribute to the bridge
emissions of the light curve. We find that 
the expected polarization characteristics are consistent with
the Crab optical data after subtraction of the DC level.
 The general features of the polarization characteristics, the light curve and 
the spectrum in optical bands have been reproduced simultaneously. 
 For the Crab pulsar, the observed position angle swing 
indicates that the viewing angle of the observer measured from the rotational
axis is  greater than $90^{\circ}$.

Although  the present model has explained the observed light
curve, spectrum and the polarization characteristics in the optical
band for the Crab pulsar, we also find that the small peak, 
which leads the main peak
and originates from the emissions inside null charge surface, becomes
to be conspicuous with increasing the energy bands if we fixed the model
parameters to produce the phase separation
$\delta\Phi\sim0.4$~phase between the main and the second peaks. 
The main peak consists of the radiations from the near the
light cylinder, while the leading peak consists  of 
the radiations near the stellar surface. The spectrum of the photons
of the leading peak becomes harder than that of the main peak, because
the magnetic field near the stellar surface is much stronger than that
near the light cylinder. The flux ratio of the leading peak
and the main peak decreases with increasing the energy bands. As a
result, the expected light curve in higher energy bands may have
triple peaks although the light curve in the optical has only two
peaks.  

\begin{figure}
\begin{center}
\includegraphics[width=15cm, height=15cm]{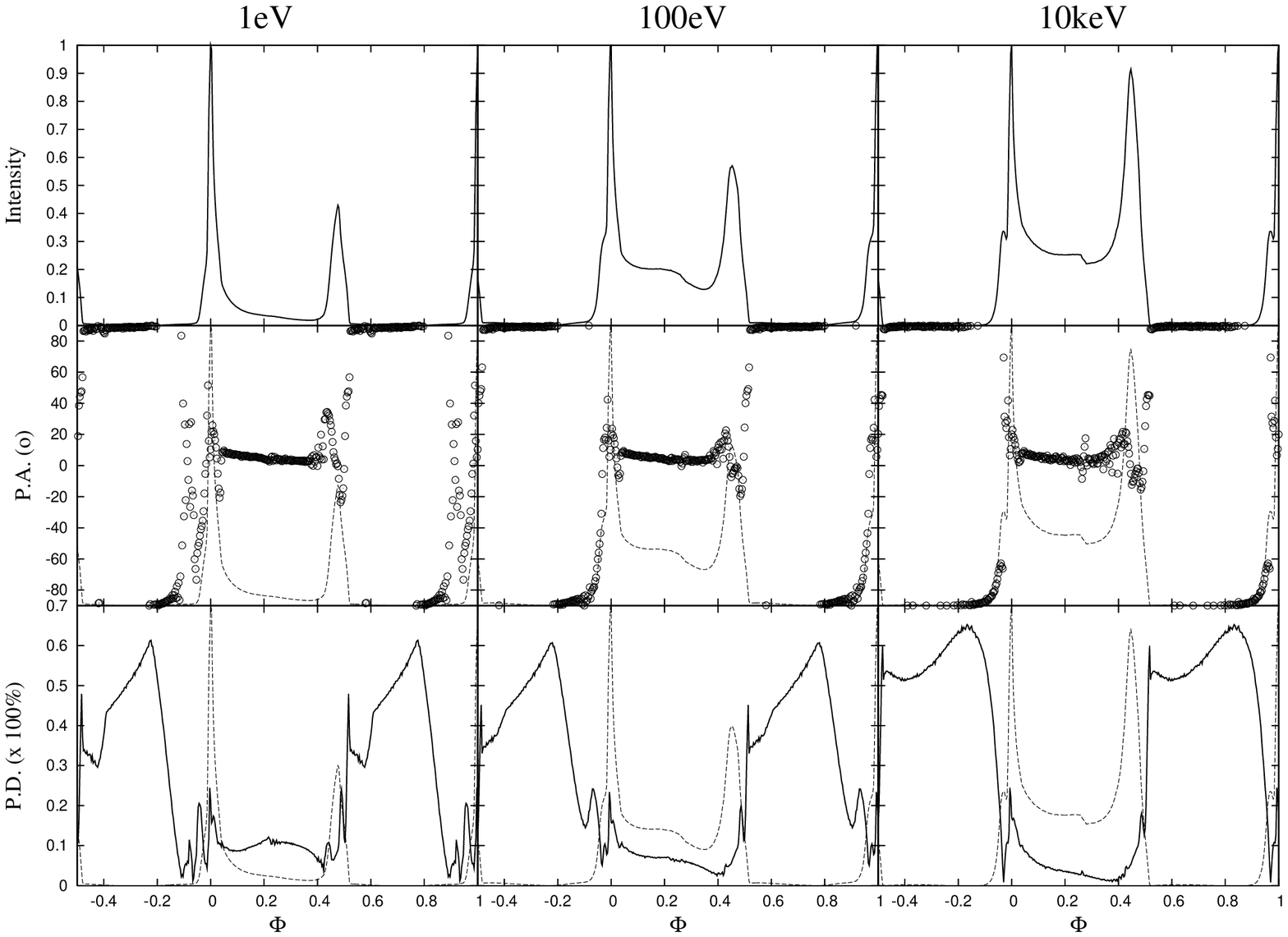}
\caption{The polarization characteristics of three energy bands, 1~eV
(left column), 100~eV (middle column) and 10~keV (right column) for
$\xi\sim95^{\circ}$, $\alpha=50^{\circ}$, $a=0.94$ and
$r_i=0.85r_n$. The present model predicts the energy dependent
polarization characteristics.}
\label{enecompari}
\end{center}
\end{figure}

If we are allowed a small discrepancy on the phase separation between
the main and the second peaks with the Crab data $\delta\Phi\sim0.4$, 
the present model can successful explain the general features
of other observed features.   
Figure~\ref{enecompari}
 summarizes the results for the viewing angles $\xi\sim95^{\circ}$ and 
$r_{i}=0.85r_{n}$ with the inclination angle $\alpha=50^{\circ}$ and
the gap upper surface of  $a=0.94$.
 We see that although the phase separation  $\delta\Phi\sim 0.45$~phase of the 
two peaks is slightly larger than $\delta\Phi\sim 0.4$~phase
 for the Crab pulsar,  the leading peak originating from the emissions
near the stellar surface keeps a low profile in the light curves in
wide energy bands. Furthermore, we note that 
the model light curves reproduce the energy dependent features of the
observed light curves that  the flux levels of the second peak 
and bridge phase increase relative to that of
the main peak, and the flux levels of the two peaks become to be the
 same at around 10~keV (Kuiper et al. 2001). 
Because the phase separation of the main and the second peaks
 depends on the configuration of the magnetic
field, the present results may predict that actual magnetic field in
the pulsar magnetosphere is modified on some level from the vacuum
dipole field, which has been assumed in the present paper, by the
plasma effects (Muslimov \& Harding 2005).

From Figure~\ref{enecompari}, we can see that the polarization characteristics
depend on the energy bands. Especially, the polarization degree in
the bridge phase decreases from about $10\%$ in optical band with
increasing the energy bands. In the optical band, the tertiary pairs
contribute to the bridge emissions with about $10\%$ of the p.d. For
higher energy bands, on the other hand,  the emissions from the
secondary pairs dominate in the bridge phase, 
because the tertiary pairs
 contribute to the bridge emissions with a smaller Lorentz factor and 
emissivity via the synchrotron process. In such a case, the
lower p.d. than $10\%$ is expected in the bridge phase as we have seen
in section~\ref{polari}. Therefore, the present
model predicts that the polarization characteristics in the bridge phase
depends on the energy bands. In the pulse and the off-pulse
phases, the polarization characteristics do not depend the energy bands very
much, because the synchrotron radiation from the secondary pairs takes a main
contribution from optical to soft X-ray emissions.  
Above hard X-ray bands, the inverse Compton process of the secondary
pairs will contribute to the emissions. Because the next generation
Compton telescope will probably be able to measure the polarization
characteristics in MeV bands, it will be required to model prediction
for it, which will be the issue in the subsequent papers. 

\acknowledgments
The authors appreciate fruitful discussion with K.Hirotani,  
S.Shibata and R.Taam. We thank L.Kuiper for providing COMPTEL,
Bepposax and OSSE data,  and the anonymous referee for his/her
helpful comments on improvements to the paper. This work was supported
by the Theoretical Institute for Advanced Research in Astrophysics
(TIARA) operated under Academia Sinica and the National Science 
Council Excellence Projects program in Taiwan administered
 through grant number NSC 94-2112-M-007-002,
NSC 94-2752-M-007-002-PAE and  NSC 95-2752-M-007-006-PAE (J.T. and
H.-K. C.), and by a RGC grant number HKU7015/05P (K.S.C.).






\clearpage

\clearpage

\clearpage

\clearpage

\clearpage

\clearpage

\clearpage

\clearpage

\clearpage
\begin{table}
\begin{tabular}{cc||c|c|c|c|c|c|c}
 & \multicolumn{1}{c}{} &
\multicolumn{7}{c}{$a=\theta_u/\theta_{lc}$} \\
 & \multicolumn{1}{c}{} & \multicolumn{1}{c|}{~~~0.9~~~} & ~~~0.91~~~ & ~~~0.92~~~ & ~~~0.93~~~ & ~~~0.94~~~ & ~~~0.95~~~ & ~~~0.96~~~ \\
\hline\hline
 & \raisebox{-1.8ex}[0pt][0pt]{$40^{\circ}$} &
\raisebox{-1.8ex}[0pt][0pt]
{$\ast$} & $\xi\sim100^{\circ}$ & $102.5^{\circ}$ & $105^{\circ}$ & $107.5^{\circ}$ & $110^{\circ}$ & $112.5^{\circ}$ \\
 &  &  & $r_{i}\sim0.75$ & 0.65 & 0.58 & 0.50 & 0.43 & 0.37 \\
\cline{2-9}
\raisebox{-1.8ex}[0pt][0pt]{$\alpha$} & \raisebox{-1.8ex}[0pt][0pt]{$50^{\circ}$} & \raisebox{-1.8ex}[0pt][0pt]{$\ast$} & \raisebox{-1.8ex}[0pt][0pt]{$\ast$} & \raisebox{-1.8ex}[0pt][0pt]{$\ast$} & $\xi\sim97^{\circ}$ & $100^{\circ}$ & $105^{\circ}$ & $107.5^{\circ}$ \\
 &  &  &  &  & $r_i\sim0.78$ & 0.67 & 0.46 & 0.37 \\
\cline{2-9}
 & \raisebox{-1.8ex}[0pt][0pt]{$60^{\circ}$} & \raisebox{-1.8ex}[0pt][0pt]{$\ast$} & \raisebox{-1.8ex}[0pt][0pt]{$\ast$} & \raisebox{-1.8ex}[0pt][0pt]{$\ast$} & \raisebox{-1.8ex}[0pt][0pt]{$\ast$} & \raisebox{-1.8ex}[0pt][0pt]{$\ast$} & $\xi\sim100^{\circ}$ & $105^{\circ}$ \\
 &  &  &  &  &  &  & $r_i\sim0.53$ & 0.30 \\
\hline
\end{tabular}
\label{table1}
\caption{Expected viewing angle and the position of the inner boundary
for various inclination angle $\alpha$ and the fraction angle of the
gap upper surface $a=\theta_u/\theta_{cl}$, where $\theta_u$ is the
polar angle of the footpoint of the magnetic surface of the gap upper
boundary and $\theta_{cl}$ is that of the magnetic surface tangent to
the light cylinder for the rotating dipole field. For each
$\alpha$ and $a$, the  viewing
angle $\xi$ is chosen to explain the phase separation of the two
peaks,  and  the position of the inner boundary $r_i$ is
determined so that
the p.a. has large swings at the both peaks and the p.d. in the off-pulse
phase becomes about $\sim60\%$. }
\end{table}
\clearpage

\end{document}